\documentclass[preprint,12pt]{elsarticle}
\usepackage{url} 
\usepackage{amsmath}
\usepackage{amssymb}
\usepackage{amsthm}
\usepackage{fontenc}
\usepackage{graphicx}
\usepackage{xcolor}
\usepackage{textcomp}
\usepackage{epstopdf}
\usepackage{braket}
\usepackage{mathtools}
\usepackage{dcolumn}
\usepackage{multirow}
\usepackage{units}
\usepackage{bbm}

\usepackage{lineno}

\begin{document}

\begin{frontmatter}

\title{Switching valley filtered current directions in multi-terminal graphene systems}
\author[uff]{V. Torres}
\author[ipnf]{D. Faria}
\author[uff]{A. Latgé \footnote{Email address: alatge@id.uff.br (Andrea Latgé ) } }
\address[uff]{Instituto de F\'isica, Universidade Federal Fluminense, Niter\'oi, Av. Litor\^{a}nea sn 24210-340, RJ-Brazil}
\address[ipnf]{Instituto Polit\'ecnico, Universidade do Estado do Rio de Janeiro, Nova Friburgo, RJ, Brazil}

\date{\today}

\begin{abstract}

Valley filtering processes have been explored in different graphene-based configurations and scenarios to control transport responses. Here we propose graphene multi-terminal set-ups properly designed to obtain valley filtered currents in a broad range of energy, besides the possibility of controlling their directions. We explore graphene systems with extended mechanical fold-like deformations as an opportunity to enhance valley filtered transmission. The mixing between the electronic confinement effects due to a magnetic field and strain results in a selective drive of the current components in the quantum Hall regime. We adopt the mode-matching method within the Green's function formalism, allowing the direct analysis of the strain effect on each valley transmission. We estimate a threshold map of confinement parameters, characterized by the magnetic, deformation, and set-up lengths, to optimize valley filter transport processes and the proper switch of the valley polarized current directions.

\end{abstract}


\end{frontmatter}

\section{Introduction}

\begin{figure}[t]
\centering
\scalebox{0.48}{
\includegraphics{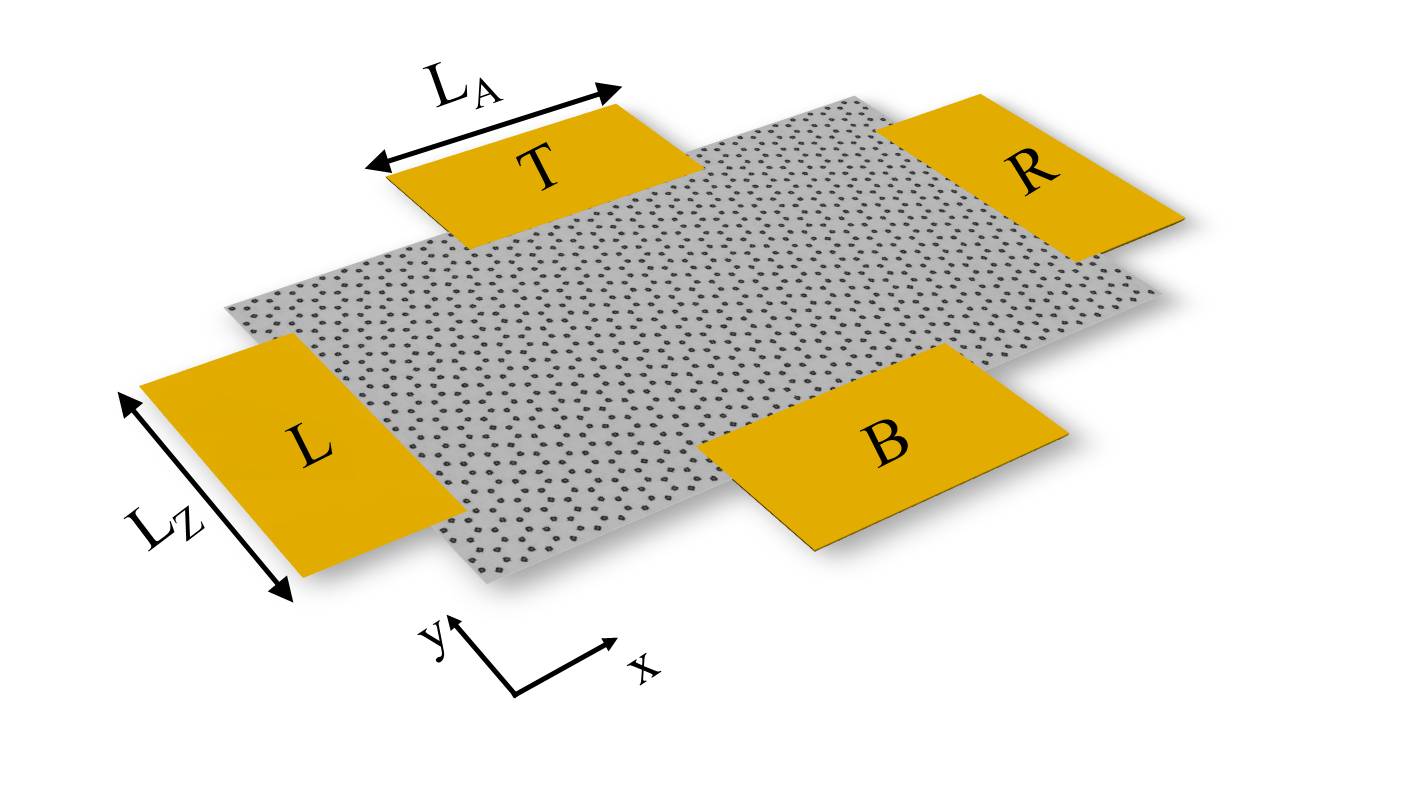}}
\caption{Schematic representation of a graphene central conductor with zigzag edges along the x-direction, contacted with semi-infinite graphene nanoribbons. Longitudinal leads, labeled as $L$ and $R$, have zigzag edges and width $L_Z$, while transverse leads, T and B,  have armchair edges and width ($L_A$).}
\label{figure1}
\end{figure}

One of the challenges for optimal use of graphene is driven by the valleytronics, in a search for taking advantage of  the extra degree of freedom given by its typical valleys K and K' \cite{yamamoto2009,schaibley2016valleytronics,settnes2017valley,rycerz2007valley,gunlycke2011graphene,gorbachev2014detecting,LINS2020353,MYOUNG2020578}. An appropriate design of the system is fundamental for controlling valley polarized currents, focusing on quantum computation applications \cite{ANG2017}. Although advances have been reported, with proposals of systems that present the valley-splitting, measurements and applications are still limited. One significant achievement in this context was the development of a 4-kink valley polarized router device based on graphene bilayers \cite{jingli2018}. Moreover, several deformed graphene systems have been explored as strategic set-ups to modulate electronic responses \cite{carrillo2016strained,Jauho2016,milovanovic2016strain,andrade2019valley,jones2017quantized,torres2019,stegmann2018current,carrillo2018enhanced,georgi2017tuning,Zhai2018}. In particular, valley-splitting occurrence and valley-inversion were proven possible experimentally in a graphene quantum dot induced by the tip of a scanning tunneling microscope (STM) in strained graphene regions \cite{freitag2018,Li2020}. The splitting of valley polarized Landau levels (LLs) caused by the coexistence of pseudo-magnetic and external magnetic fields was observed in strained graphene in the quantum Hall regime \cite{Li2020,lili2015,klimov2012electromechanical}. These measurements motivate further research on graphene strained systems with specific designs enabling the collection of valley filtered currents.      

Valley filtered transport has been explored in different scenarios. Valley polarized currents were proposed based on a ballistic point contact with zigzag edges allowing polarity inversion by local application of a gate voltage \cite{rycerz2007valley}. Also, strain in graphene is known to raise the possibility of valley spatial-separation. Depending on the deformation profile, it is possible to generate valley polarized local density of states (LDOS) \cite{settnes2017valley, Jauho2016} in regions that work as wave-guides for polarized currents  \cite{carrillo2016strained,torres2018tuning,faria2020valley}. Other works have explored valley filtered currents in graphene considering the valley spatial-separation combined with different mechanisms, such as edge disorder, strain superlattices, external magnetic fields, and multi-terminal configurations \cite{carrillo2016strained,milovanovic2016strain,torres2019, Fujita2018}. For example, graphene superlattices designed by out-of-plane Gaussian deformations are shown to improve
the valley filter capabilities of a single perturbation, with the conductance exhibiting a sequence of valley filtered plateaux \cite{torres2019}.

External magnetic fields are used as an alternative mechanism to enhance the valley spatial-separation and valley filtering effects due to strain \cite{Li2020,lili2015,faria2020valley}. The external field introduces the time-reversal symmetry breaking allowing the manipulation of valley-currents in different directions. Additionally, since in the Quantum Hall regime the energy levels are known for pristine systems, the effect of deformations becomes more evident \cite{settnes2017valley,milovanovic2016strain,low2010,Rainis2011,kim2011,Qi2013,Roy2013,Farajollahpour2017}. The detection of valley polarized LLs has been predicted recently, in the LDOS calculation of a fold-like deformed graphene \cite{faria2020valley}. The spatial evolution of the valley-dependent features is revealed by a braided structure correlated with the pseudo-magnetic field fringes arising on strained graphene that may be observable in STM measurements. In the two-lead device, extra conducting channels were predicted due to the deformation, expected to be valley polarized. However, further analysis should be done to collect the polarized states in multi-terminal geometries.  

Here we explore different possibilities of filtering valley-currents by conveniently attaching electronic contacts where the current may go through. The idea is to provide appropriate conditions to separate and detect valley polarized transport. We consider a 4-terminal strained fold graphene system, with leads positioned transversely and longitudinally to the central conductor. Some states are confined in the deformation region, while other polarized states are pushed to the transverse leads, with valley filtered currents expected in a broader energy range depending on the strain intensity. We show that by bringing the system into the quantum Hall regime, it is possible to switch the polarized current directions with filtered-transport in the longitudinal terminals. We discuss the relevance of the system's different confinement parameters to obtain valley filter transport, and valley polarized current directions switch.

\section{Model}

We consider a system formed by a graphene central conductor with zigzag edges along the $x$-direction, connected to terminals that are described by perfect semi-infinite nanoribbons, labeled as $L$ (left), $R$ (right), $T$ (top), and $B$ (bottom). The contact widths with zigzag and armchair edges are $L_Z$ and $L_A$, respectively, as shown in Fig. \ref{figure1}. The system is modeled by the first-neighbor tight-binding Hamiltonian, given by

\begin{eqnarray}
H=\sum_{\langle i,j\rangle} \lambda_{ij}c^{\dagger}_{i}c_{j}+\sum_{l=1}^{4}\left(\sum_{\langle i,j\rangle} \lambda_{ij}^{l}c^{\dagger l}_{i}c^{l}_{j} +\sum_{\langle i,j\rangle} h^{l}_{i,j}c^{\dagger l}_{i}c^{l}_{j} \right)\nonumber\\
\end{eqnarray}

\noindent where the first term describes the conductor's central region and the second refers to the 4-terminals, which are coupled with the central part by the hopping energy $h^{l}_{i,j}$. The modification in the hopping parameter due to mechanical deformations and an external magnetic field, applied perpendicularly to the system, is \cite{castro2009electronic}

\begin{equation}
\begin{split}
\lambda_{ij}=\lambda_{0}e^{-\beta\left(\frac{l_{ij}}{a_c}-1\right)}e^{i \frac{2\pi e }{h}\Phi_{ij}}, 
\end{split}
\end{equation}

\noindent where $\lambda_0=2.75eV$, $\beta\approx 3$, $a_c=1.42\AA$ is the carbon-carbon distance in the unstrained system, and  $l_{ij}=\frac{1}{a}\left(a^2+\varepsilon_{xx}x^2_{ij}+\varepsilon_{yy}y^2_{ij}+2\varepsilon_{xy}x_{ij}y_{ij}\right)$ is the new distance between the carbon atomic sites $i$ and $j$, written in terms of the strain tensor $\varepsilon_{\mu\nu}=1/2\left(\partial_\mu u_\nu +\partial_\nu u_\mu+\partial_\mu h \partial_\nu h\right)$ as a function of in-plane, $u_{\mu(\nu)}$, and out-plane, $h$, deformations, where $\mu$ and $\nu$ are $x$ and $y$ directions.  The deformation considered extends from left to right contacts, along the zigzag direction, while top and bottom leads are considered as pristine armchair nanoribbons.  The effect of a magnetic field $\boldsymbol{B}$, is introduced in the tight-binding Hamiltonian via the Peierls' approximation \cite{low2010}, where the phase factor $\Phi_{ij}$ depends on the potential vector $\boldsymbol{A}$. The gauge was conveniently chosen to preserve the periodicity in the four terminals  \cite{power2017electron}. 

To explore the valley-currents, we consider a fold-like deformation, representative of extended deformations usually found in graphene samples \cite{Li2020,lili2015,Yamamoto2012,Tapaszto2012,Calado2012,Lim2015,Jiang2017,Ma2018,Carbone2019}, defined as \cite{carrillo2016strained,faria2020valley,torres2017gap}
\begin{equation}
\begin{split}
h=A_{f} e^{ -\frac{(y-y_0)^2}{b^2}},
\end{split}
\end{equation}

\noindent where $A_f$ and $b$ denote the fold amplitude and extension, respectively, and $y_0=L_Z/2$ corresponds to the deformation center. The parameter considered to indicate the deformation intensity is $\alpha=A_f^2/b^2$, which corresponds to a maximum strain intensity $\varepsilon_M=\alpha/e$, where $e$ is the Euler's number.  Modifications in the hopping parameter due to the mechanical deformation give origin to a pseudo-gauge field in the continuum description \cite{vozmediano2010gauge,katsnelson2007graphene,de2013gauge}, written as

\begin{equation}
\begin{split}
\boldsymbol{A}_{ps}=\frac{\beta\hbar v_f}{2a_{c}}\left(\varepsilon_{xx}-\varepsilon_{yy},
-2\varepsilon_{xy} \right)\,\,,
\end{split}
\end{equation}

\noindent with $v_f$ being the Fermi velocity. The pseudo-magnetic field $\boldsymbol{B}_{ps}=\boldsymbol\nabla \times \boldsymbol{A}_{ps}$ for this deformation exhibits a characteristic strip pattern that alternates between positive and negative field regions. It has opposite signs for electrons around each valley, allowing the production of valley-current polarization on the pseudo-field stripes \cite{carrillo2016strained}.

A continuum model analysis for the two-dimensional graphene shows that in the presence of an external magnetic field, fold-like deformations generate new states within the LLs, which can be characterized by $\gamma=l_B/b$, the ratio between the magnetic length, $l_B=\sqrt{\hbar/eB}$, and the deformation width, $b$ \cite{faria2020valley}. This parameter can be used as a guide to define different regimes of valley filter realizations. In the system considered here, since the central conductor is finite, with the same width as the zigzag terminal leads, we will show the relevance of a third parameter related to the deformation, the ratio between the central conductor and the deformation widths, $L_Z/b$, called as the strain spread in the system.  

We adopt the Mode-Matching Method \cite{ando1991quantum,khomyakov2005conductance,sorensen2009efficient} that allows a direct analysis of the deformation effects in each valley. The transmission matrix elements for one electron in the mode $n$ coming from the $p^{th}$- terminal, scattered to a mode $m$ in the $l^{th}$- terminal, is written as \cite{khomyakov2005conductance}
\begin{equation}
\begin{split}
&
t^{k,m,n}_{l,p}=\sqrt{\frac{v^{k}_{l,n}}{v^{k}_{p,m}}}\left[({u}^{k}_{l,n})^{\dagger}G(E)\left[ G^{0}(E)\right]^{-1}u^{k}_{p,m}\right],
\end{split}
\end{equation}

\noindent where ${u}^{k}_{l,n}$ is the terminal eigenvector in the propagating mode $n$, $v^{k}_{l,n}$ is the Bloch velocity for the $n^{th}$- mode, and $k$ is the valley index ($k=K,K'$). The Green's functions of the full system and the terminals, $G(E)$ and $G^{0}(E)$, respectively, are obtained by standard iterative techniques \cite{lewenkopf2013recursive,thorgilsson2014recursive}. 

The total valley-dependent transmission is defined as
\begin{equation}
\begin{split}
\tau^{k}_{l,p}=\sum_{m,n}\mid t_{l,p}^{k,n,m}\mid^2\,\,,
\end{split}
\end{equation}
\noindent that allows the total transmission calculation \cite{datta1997electronic}, $T_{l,p}=\sum_{k}\mid \tau^{k}_{l,p}\mid^2$.

\section{Valley-transport in the strained system} 

\begin{figure}[ht!]
\centering
\scalebox{0.52}{
\includegraphics{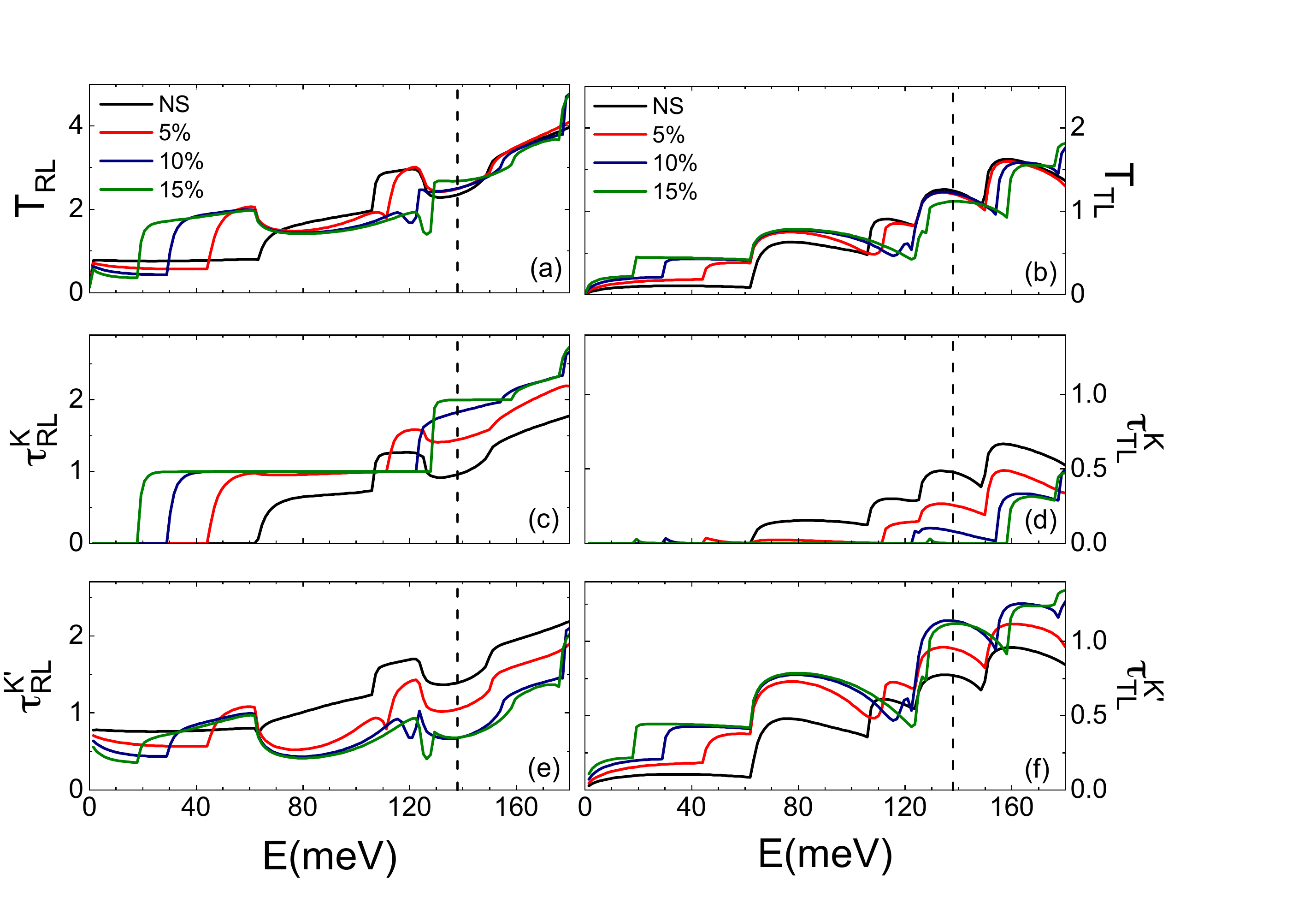}}
\caption{Total and valley-dependent transmission coefficients between left and right terminals (a,c,e) and left and top (b,d,f) leads for the unstrained system (NS) and increasing strain intensities with $\alpha= 5$, $10$, and $15 \%$. 
Parameters: $ L_Z = 42.5nm$, $ L_A =29.3nm $, and $ b = 30 {a_c}$.}
\label{figure2}
\end{figure}


We analyze first the effect of the extended deformation on the transport properties of the multi-terminal system, focusing on the valley-dependent transport responses. 
The total transmission results, $T_{RL}$ and $T_{TL}$, are shown in Fig.\ \ref{figure2} (a) and (b), respectively, for fold structures with no strain (NS) and different maximum strain intensities.  Due to the system symmetry, the transmissions $T_{TL}$ and $T_{BL}$ are the same, as expected. Valley-dependent transmission components $T_{RL}$ and $T_{TL}$ are depicted in Figs.\ \ref{figure2} (c) and (d) and Figs.\ \ref{figure2}  (e) and (f) for valleys $K$ and $K'$, respectively. Note that for the undeformed 4-terminal devices (black curves), for energies below approximately $65 meV$ , only the edges states corresponding to $K'$-valley contribute to the left to right (left panels) and left to top (right panels) transmissions. This feature is expected for the first plateau in the zigzag nanoribbon leads. Then, the four-terminal system's geometry allows the filtering of $K'$-valley electrons corresponding to the edges states of the zigzag leads. When the deformation is introduced, the mechanical perturbation induces other plateaux formation at lower energies [Fig.\ \ref{figure2} (a) and (b)]. The left to top transport is still formed essentially by the $K'$-valley electrons, which generates a valley polarized current in larger energy ranges depending on the strain intensity. As the deformation increases, the enhancement of the valley- filtering process in the left-top transmission [Fig.\ \ref{figure2} (d) and (f)] is followed by an increase of the left to right transmission [Fig.\ \ref{figure2} (c) and (e)]. But the longitudinal transmission $T_{RL}$ is given by a combination of both valley contributions, except for a small energy range, next to zero.  

\begin{figure}[ht!]
\centering
\scalebox{1.}{
\includegraphics{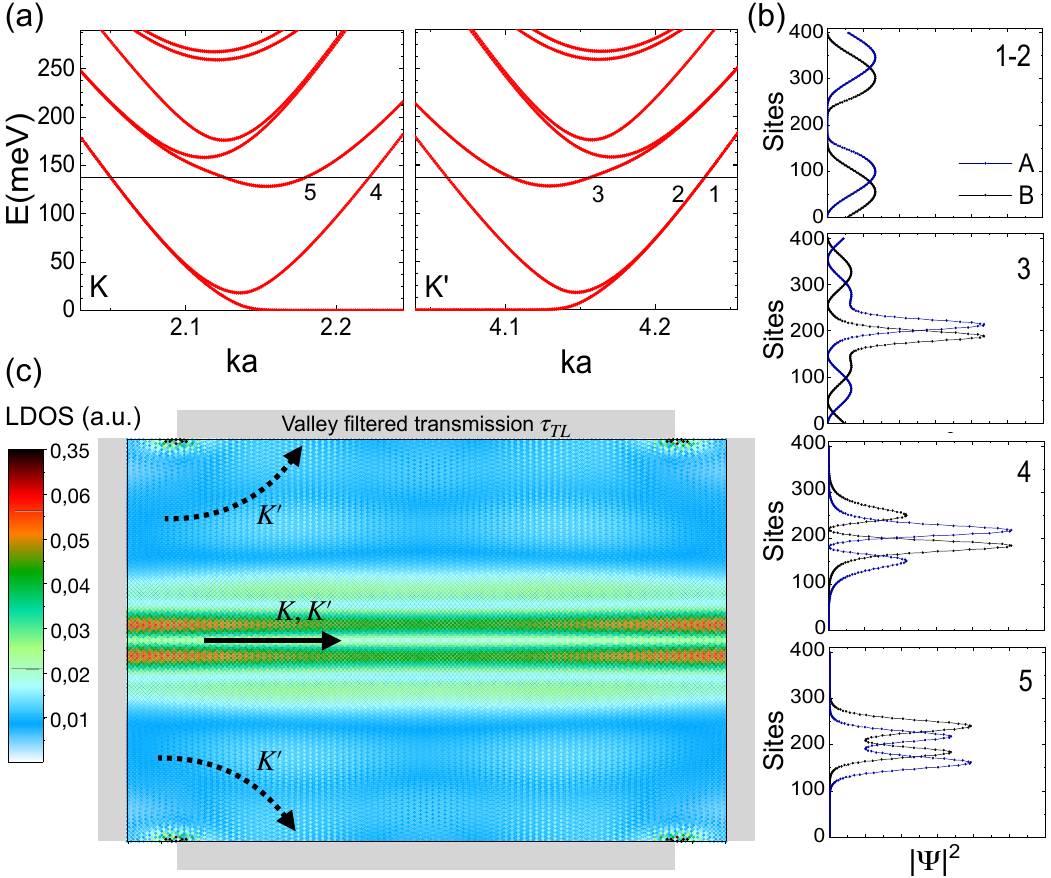}}
\caption{(a) Band structure of a folded zigzag nanoribbon. (b) Probability density of states at $E=138 meV$ corresponding to the momenta labeled as 1, 2, and 3 ($K'$ valley), and  4 and 5 (valley $K$) in (a). Blue and black symbols indicate the probability densities of $A$ and $B$ sub-lattices, respectively. (c) LDOS of the central part of an equivalent 4-leads system, at the same energy as pointed in (b). Schematic black dashed arrows in (c) indicate valley filtered left to top (and bottom) transmission. The black arrow pointing from left to right corresponds to non-polarized transmission in this direction. Parameters: $L_Z=42.5nm$, $L_A=29.3nm$, $b = 30{a_c}$ $A_f = 11.6{a_c}$, and $\alpha=15\%$}
\label{figR02}
\end{figure}

To better characterize the deformation effects, taking into account the fact that electrons are injected from the left terminal, we show in Fig.\ \ref{figR02} (a) the electronic band structure of a zigzag nanoribbon with the same width $L_Z$ as the central part of the 4-leads system. Fig.\ \ref{figR02} (b) illustrates the spatial distribution of the probability density of the states that contribute for a particular energy value equal to $E=138 meV$, marked in Fig.\ \ref{figR02} (a). Differently from the results of the unperturbed system where the states are spread along all the nanoribbon (not shown), in the deformed system, a high electronic concentration is observed at the deformation region (note in Fig.\ \ref{figR02} (b), the state labeled as 3 at $K' $-valley, and 4 and 5 at valley $K$, with positive velocities). Contrarily, the other electronic states 1 and 2, coming from the $K'$- valley, are localized at the ribbon edges. These states contribute to the valley polarized transmission from left to top/bottom leads when transverse leads are connected to the central system in the 4-lead configuration. The confinement introduced by the extended deformation is easier to understand as a consequence of the pseudo-magnetic field. The current injected along the deformed system is expected to be divided into two main contributions, one along the center of the deformation with a high density of states, formed by states from both valleys, and the other contribution, at the fold tails, is formed by the states from valley $K’$. In the 4-leads configuration, the top and bottom contacts collect these states to the leads, making the filter process feasible at higher energies. These features are summarized in LDOS for the 4-lead configuration, shown in Fig.\ \ref{figR02} (c), where the schematic arrows highlight the discussed valley selective transport in the system. 

As discussed, the addition of new contacts (top and bottom), combined with the deformation effect, creates a favorable scenario for getting valley polarized currents. Previous studies, however, demonstrated that the electronic transport is affected by edge roughness in nanoribbons \cite{mucciolo2009conductance,faria2020valley}. To eliminate possible effects of edge disorder, we propose applying a magnetic field in the system to achieve suitable conditions for valley filter transport, considering much smaller strain intensities. As we show next, the states propagating in the deformation's central region give origin to the polarized current is originated, with the valley filter happening in the left-right current direction.


\section{Switching the valley filter direction with a magnetic field} 
 
\begin{figure}[ht!]
\centering
\scalebox{0.76}{
\includegraphics{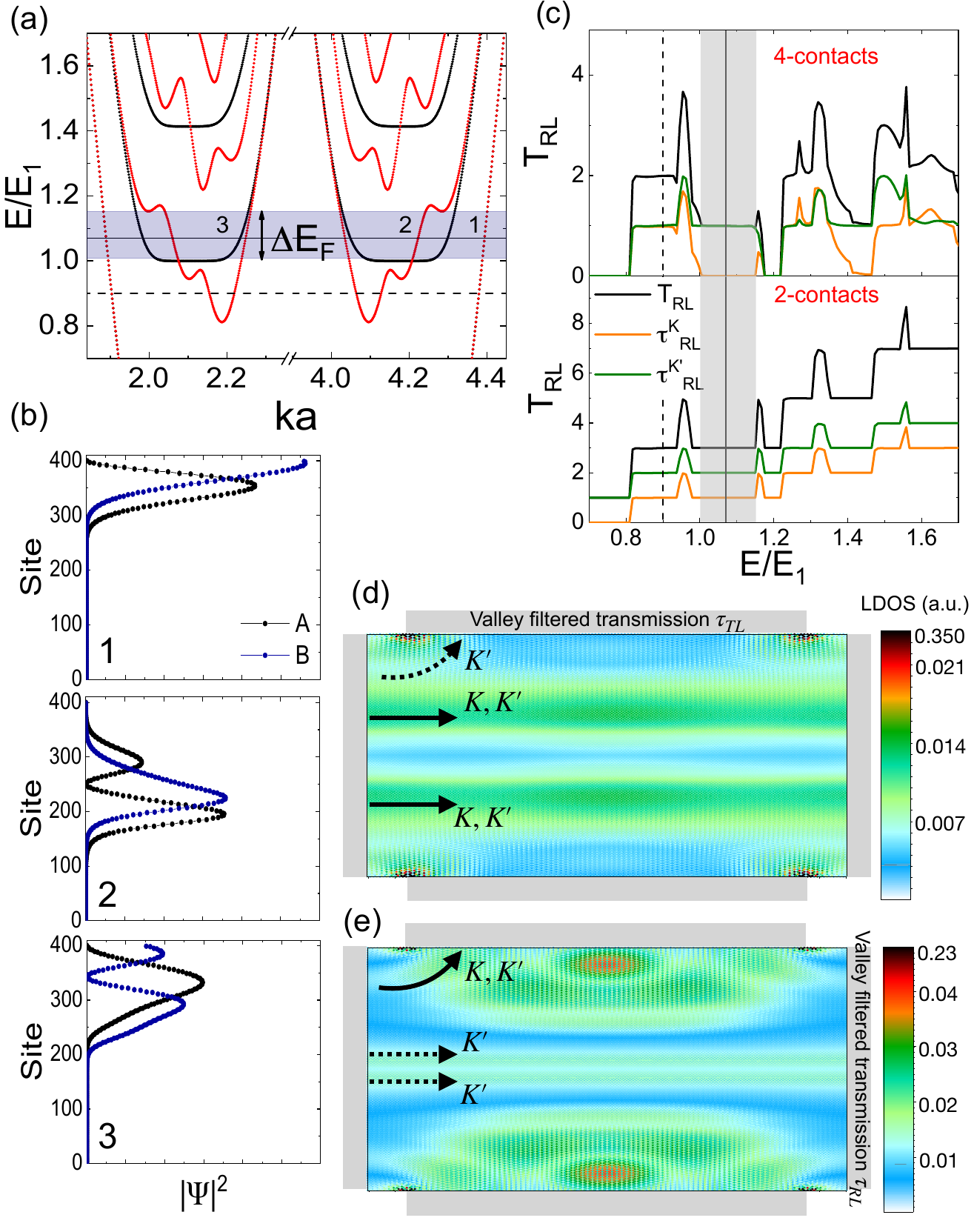}}
\caption{(a) Band structure of a strained (red) and unstrained (black) zigzag nanoribbon with $B=25T$. (b) Probability density at energy $ E/E_1 = 1.07$ for the momentum states labeled as 1 and 2 (valley $K'$) and 3 (valley $K$). Blue and black curves indicate the probability densities of $A$ and $B$ sub-lattices, respectively. (c) Comparison of transmission coefficients from left to right terminals $T_{RL}$ for deformed systems with 4 and 2 contacts and $B=25T$. Green and orange curves indicate the transmission of the valleys $K'$ and $K$, respectively. LDOS for the 4-leads system at energy (d) $E/E_1=0.90$ and (e) $E/E_1=1.07$. Dashed (continuous) arrows indicate valley filtered (non-filtered) transmission. Parameters: $L_Z=42.5nm$, $L_A=29.3nm$, $ b = 40 {a_c} $, $ A_f=8.9a_c $.}
\label{fig03}
\end{figure}

To guarantee the formation of the LLs,  magnetic field values of the order of 25T have been considered at first. However, as we will show this analysis works equally well for experimentally feasible magnetic field values.

Results for the electronic band structure and the probability density of the states contributing at energy $E/E_1=1.1$ are presented in Fig.\ \ref{fig03}, for the same zigzag nanoribbon discussed in Fig.\ \ref{figR02}, now under the effect of a magnetic field B=25T, for a smaller strain intensity.  The energies are given in terms of the first Landau state, $E_1=3\lambda_0 a_c\sqrt{2}/2l_B$. For comparison, Fig.\ \ref{fig03} (a) shows the system band structure without mechanical deformation indicated by the black curve. The inhomogeneous pseudo-magnetic field forms new dispersive states, characterized by states with maximum and minimum energies for each level. 
The probability densities for the momentum states labeled as 1, 2, and 3 are depicted in Figs.\ \ref{fig03} (b), for both sub-lattices (black and blue symbols). The states 1 (valley $K'$) and 3 (valley $K$) are localized closer to the top edge. These states would be driven to a top contact if connected. Reversing the magnetic field direction should move the states to the bottom contact. Otherwise, state 2 comes from the valley $K'$, localized in the ribbon's center. This state will be responsible for a $K'$-filter phenomena observable in the left-right transmission $T_{RL}$, for this particular energy. This filter takes place not only at that energy but in an energy range labeled as  $\Delta E_F$ [light purple shaded strip in Fig.\ \ref{fig03} (a)]. The filter range is revealed in the transmission results shown in Fig.\ \ref{fig03} (c) 4-contacts device (light purple shaded strip), with only the $K'$-valley contribution, in contrast to the results for the 2-contact set-up (nanoribbon) where the two valleys contribute to the transmission.
 
The LDOS maps in the 4-lead system help to identify the spatial electronic distribution, as shown Fig.\ \ref{fig03} (d) and (e) at energies $E/E_1=0.90$ and $1.07$, respectively. 
The LDOS at energy $E/E_1=0.90$ shows some electronic concentration at the central part of the system, also exhibiting higher electronic distribution closer to the interface with the top and bottom leads due to polarized edge states that flow to top leads. In contrast, at $E/E_1=1.07$, a higher electronic concentration is noticeable at top and bottom contact entrances, highlighted with orange-colored LDOS. In the central part of the system, some localization is still present. 
One should take into account that in contrast to the information obtained from the probability density of the individual k-states, discussed in Fig.\ \ref{fig03} (b), the LDOS counts the full contribution at a particular energy, which includes the probability density of states with both velocity directions. The selected velocity direction defines the electronic carrier flux. In the present case, choosing the electron departing from the left terminal, the magnetic field pushes the carrier to the top lead unless it gets trapped in the deformation due to the pseudo-magnetic field confinement. An analysis of the valley-dependent transmission indicates a mixed contribution of valleys K and K' in the top contact. Simultaneously, in the left-right direction, polarized carriers are present, as will be discussed next. To investigate the transport through the individual leads and the filter process's dependence on the magnetic field intensity, we calculate the transmission components, considering an electron flowing from the left terminal. 
We discuss the competition between the localization mechanisms introduced by strain and magnetic field, in terms of the parameter $\gamma=l_B/b$. 


\begin{figure}[ht!]
\centering
\scalebox{0.5}{
\includegraphics{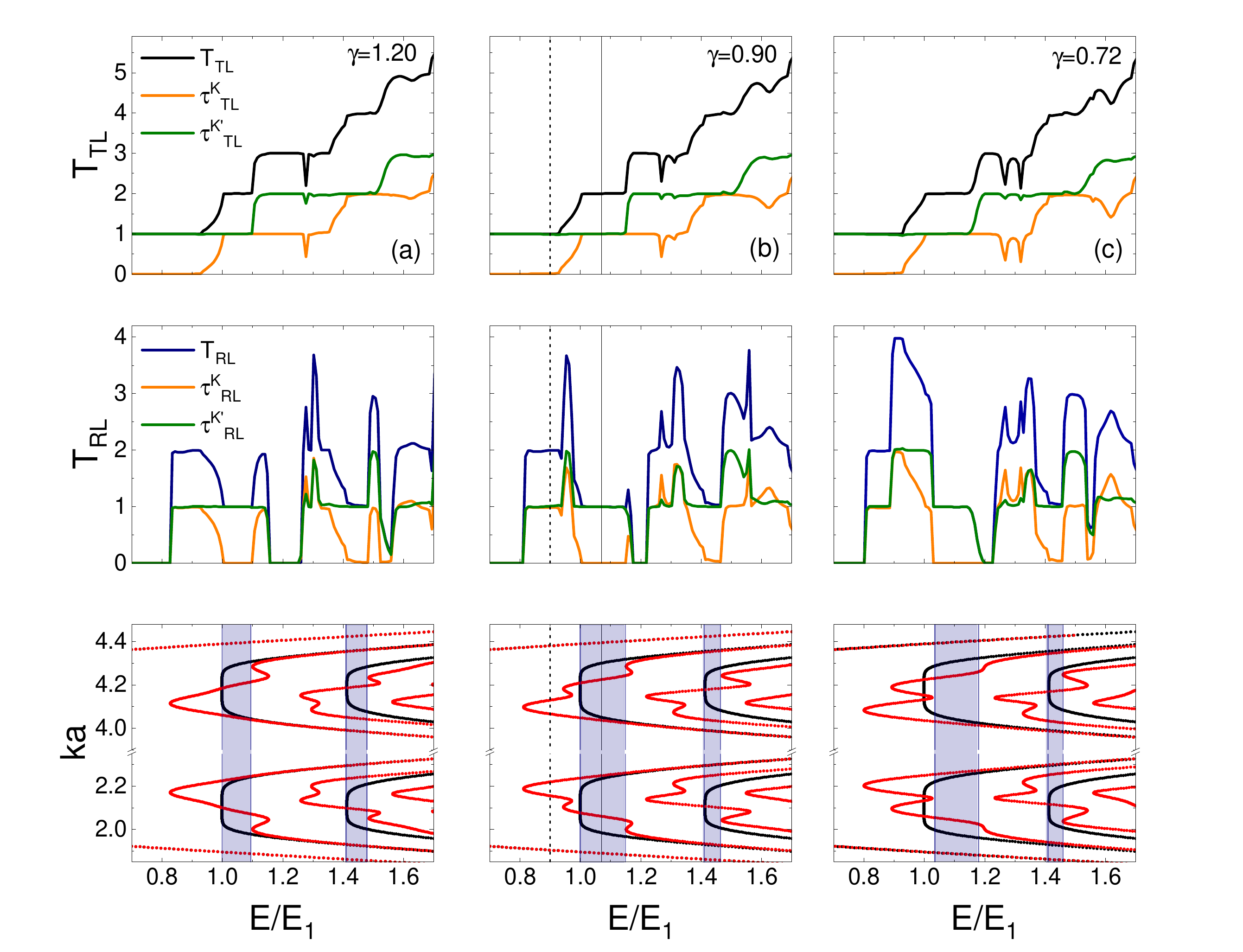}}
\caption{Top Panels: Transmission coefficients from left to top terminals $T_{LT}$ (first row) and left to right $T_{LR}$ (second row) for a fold zigzag nanoribbon with $B=25T$ and $\gamma$ equal to (a) $1.20$ (b) $0.90$  and (c) $0.72$. The orange and green curves indicate transmission from $K$ and $K'$ valley, respectively. Parameters: $L_Z=42.5nm$, $L_A=29.3nm$, $ \alpha = 5\%$ and (a) $ b = 30 {a_c} $, (b) $ b = 40 {a_c} $,  (c)$ b = 50 {a_c} $. Bottom Panels: Corresponding band structures for the strained (red curves) and unstrained (black curves) zigzag ribbons.}
\label{figure5}
\end{figure}


The transmission results for the 4-lead system with different $\gamma$ values are shown in Fig.\  \ref{figure5}. Differently from the case of zero field, the transmission from left to bottom leads (not shown) goes to zero, $T_{BL}=0$, due to the Lorentz force. On the other hand, the transmission from left to top, $T_{TL}$, is formed by a combination of carriers from both valleys $K$ and $K'$, as can be seen in top panels, first row,  in Fig.\ \ref{figure5} (a), (b) and (c) (orange and green curves).
The left to right transmission results, presented in  the top panels, second row, show that only valley $K'$ contributes in some energy windows. This feature confirms that it is possible to switch the valley filter to the left-right direction when the magnetic field is turned on. We show next further analysis to identify the filtering energy window dependence on the system's parameters. 

As mentioned, the filter region is closely related to the features of the zigzag nanoribbon terminals' electronic structure.  We observe in the case of $\gamma=1.20$ and $\gamma=0.90$ that $\Delta E_F$ is bounded between the first LL energy and the most external minimum of the energy band, which is indicated by the shaded light purple stripe in the bottom panel of Fig.\ \ref{figure5} (a) and (b). For smaller $\gamma$ parameters, new states of maximum and minimum energies are formed in the band structure, overcoming the first LL.  This feature is seen in the case of  $\gamma = 0.72$ where $\Delta E_F$  is now between a maximum and the minimum of energy [bottom panel of Fig.\ \ref{figure5} (c)].

\section{Valley filter windows: dependence on the confinement parameters} 

\begin{figure}[ht!]
\begin{center}
\scalebox{0.67}{\includegraphics{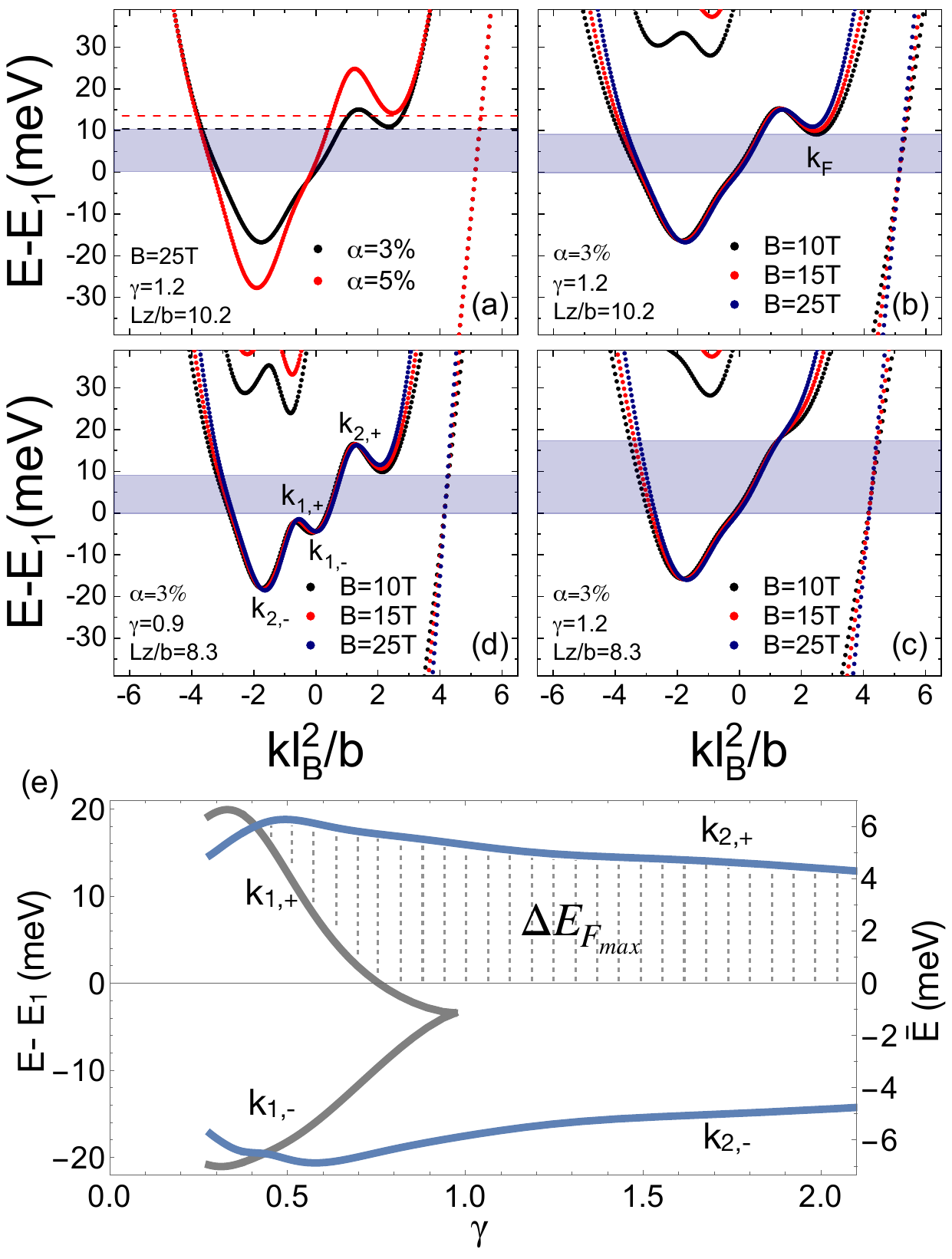}}
\caption{Band structure of fold zigzag nanoribbons with energy given with respect to the first LL of the unperturbed system. (a)B=25T, $\gamma=1.2$, $L_Z/b=10.2$ and strain intensities $\alpha=3\%$ and $5\%$, depicted in black and red curves, respectively. (b) and (c) comparison between different values of $B= 10, 15$ and $25T$, for $L_Z/b=10.2$ and $8.3$, respectively, with $\alpha=3\%$ and $\gamma=1.2$. (d) Comparison between the same magnetic field values, with $\gamma=0.9$, $L_Z/b=8.3$ and $\alpha=3\%$. Extreme energy variations are labeled by the momentum values $k_F$, $k_{1,\pm}$ and $k_{2,\pm}$. (e) Energy filter window as a function of $\gamma$, defined by the extreme energy corrections given by the momentum values labeled as in (d). 
\label{bandask2}}
\end{center}
\end{figure} 

We propose an energy filter route to predict the valley filter window in terms of controllable parameters given by the magnetic field, strain intensity, and ribbon widths. We alternatively propose a description of the strain effect by the amount of deformation spread in the ribbon width, given by $L_Z/b$; big ratio pointing to the weakly spread of the system's deformation.
The energy bands of fold zigzag nanoribbons for different combinations of $\gamma$ and $L_Z/b$, and magnetic field intensities are shown in Fig.\ \ref{bandask2}. The carrier's energies are defined with respect to the first LL of the unstrained system $(E-E_1)$. Also, to better visualize the magnetic field effect on the electronic structure of unstrained nanoribbons, an effective dimensionless momentum is usually adopted \cite{Delplace2010}, $kl_B^2/L_Z$. Here, we propose a similar renormalized momentum, given in terms of the deformation extension, $kl_B^2/b$.   
 
To highlight the strain effect through the new parameter $L_Z/b$, we present in Fig.\ \ref{bandask2} (a)-(d) band structure results, taking fixed values of this ratio in each panel. In Fig.\ \ref{bandask2}(a) we compare the predictions obtained by longitudinal transmission results for the valley filter window $\Delta E_F$ at fixed magnetic field value $B=25 T$, for the strains $\alpha=3\%$ and $5\%$, marked by the light purple shaded strip and red dashed line, respectively. The results reveal an increased filtering energy window for higher $\alpha$ intensity.
Within the proposed momentum scale, changes in the magnetic field intensities ($B=10$, $15$, and $25$ T) slightly modify the energy filter range, as can be seen in Fig.\ \ref{bandask2} (b) to (d). The light purple shaded strips in panels (b-d) correspond to the valley filtering windows for the left to right transmission coefficients obtained with $B=10T$. Additionally, the momentum $k_F$, marked in panel (b), corresponds to the minimum energy value limiting the valley filter region. 

Comparing the band structures depicted in Figs.\ \ref{bandask2} (b) and (c), and the corresponding filter energies for two different values of $L_Z/b$, we conclude that a reduction of $L_Z/ b$ from $10.2$ to $8.3$ affects the edges states, pushing the states positioned at the deformation tails to the top contact. The valley filter window efficiency in the longitudinal direction is then enhanced in a larger window. Otherwise, a direct comparison between Figs.\ \ref{bandask2}(c) and (d) for fixed $L_Z/b$ ratio equal to 8.3, indicates that the filter region is larger for $\gamma=1.2$  than for $\gamma=0.9$. Smaller $\gamma$ implies electrons more confined in the deformation region, decreasing the filter window efficiency. 
  
We obtain the maximum valley filter energy window by mapping the extreme energy values (maxima and minima) in the zigzag carrier's band structure. The states with extreme energy values are labeled as $k_{1,\pm}$ and $k_{2,\pm}$, as shown in Fig.\ \ref{bandask2} (d).  
The evolution of the extreme energies as a function of $\gamma$ is presented in Fig.\ \ref{bandask2} (e) for $\alpha=3\%$. Alternatively, the  the right axes is scaled as $\bar E= (E-E_1)/\alpha$. This energy scale is possible due to the linear dependence of the energy variation on the strain $\alpha$, as predicted by perturbation theory \cite{faria2020valley}. A small asymmetry of the energy variation concerning the first LL (zero horizontal line) is found, being more noticeable for the $k_{1,\pm}$ states. The states $k_{1,\pm}$ rise as energy extrema for $\gamma < 1$, with energy correction lower than the first LL. For values of $\gamma<0.4$, the energy of the state $k_{1,+}$ is larger than the first LL. Comparison of these results with longitudinal transmission calculations allowed the identification of the maximum valley filter energy window $(\Delta E_{F_\text{max}})$, highlighted with grid lines in the figure. We find two different regimes; for $\gamma > 0.75$, the filter window is given by the energy difference between states $k_{2,+}$ and the first LL energy.  For $\gamma < 0.75$, energy deviations for states $k_{1,+}$ are larger than zero, then the valley filter window is given by the energy difference between the states $k_{1,+}$ and $k_{2,+}$. This energy difference starts to reduce for $\gamma$ smaller than $0.75$ and around $\gamma= 0.4$ is closed, indicating a lower boundary for valley filtering energy processes.

\begin{figure}[ht!]
\begin{center}
\scalebox{0.38}{\includegraphics{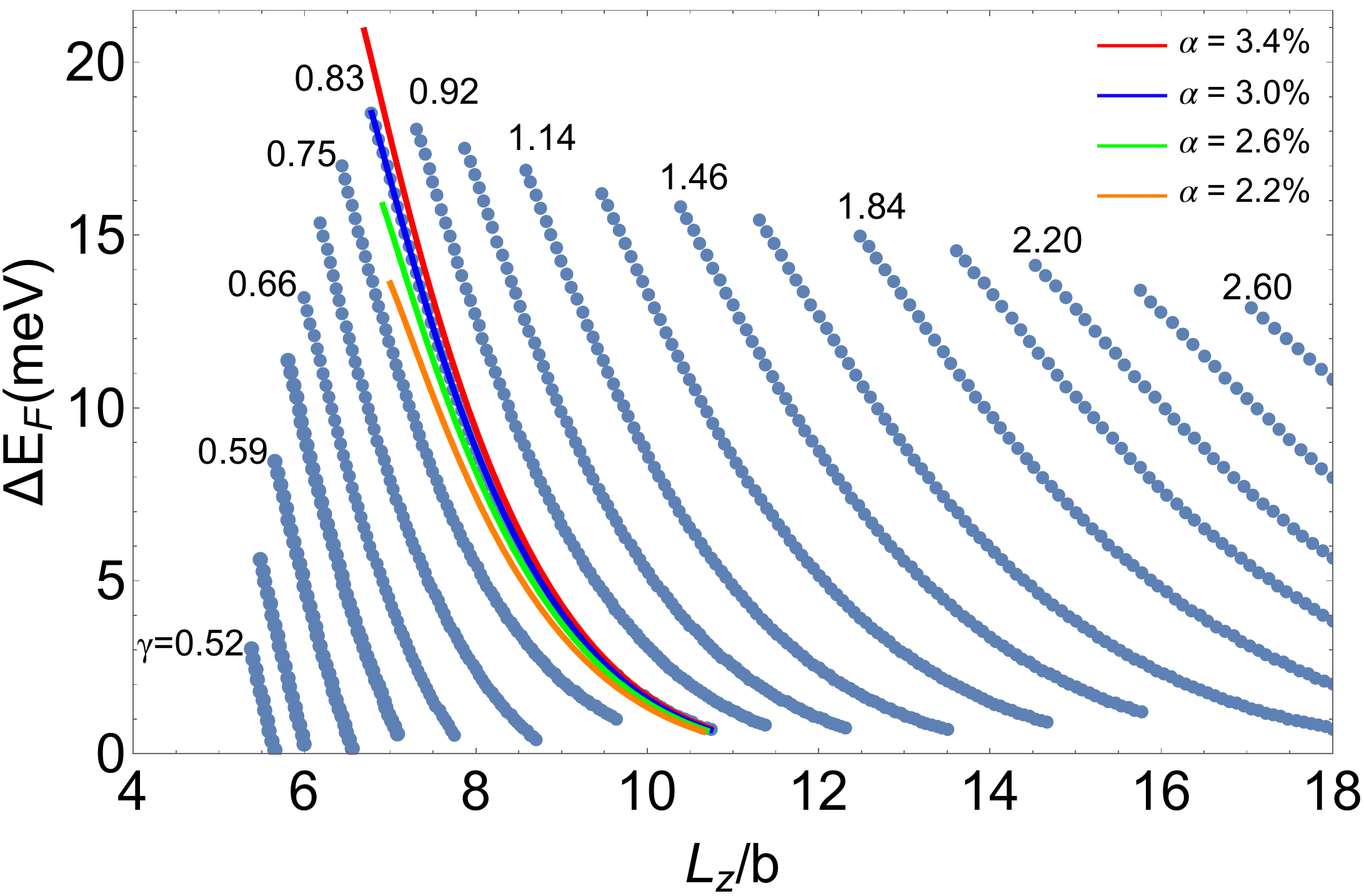}}
\caption{Energy dependence of the $K'$-valley filtered left to right transmission on the 4-leads system confinement parameters, $\gamma$, and $L_Z/b$, for a fixed magnetic field $B=10 T$ and strain intensity $\alpha=3.0\%$ (blue curves). Valley filtering energy window dependence on $\alpha$ values (red, green, and orange lines) for $\gamma=0.83$.
\label{filtermap}}
\end{center}
\end{figure} 

Finally, we show in  Fig.\ \ref{filtermap} the  
valley filter energy window dependence on the strain ratio $L_Z/b$, for different $\gamma$ values, B=10T, and $\alpha=3.0\%$.
The individual curves correspond to a fixed $\gamma$ ratio. 
Notice that values of $\gamma$ around $0.8$ have larger filter energy windows. Moreover, there will be a minimum value for $\gamma$ at which the valley filter goes to zero, as discussed previously.  
Meanwhile, the spread of strain in the system, $L_Z/b$, was also shown to be extremely relevant for enhancing the valley filtering energy window. The evolution of the curves maps an energy-topography diagram that defines ideal parameter ranges ($\gamma$ x $L_Z/b$) for the occurrence of valley filtering. For example, to achieve valley filtered in the left to the right transmission for $\gamma=0.83$, the $L_Z/b$ range should be between $7.5$ and $10.5$. Also shown in Fig.\ \ref{filtermap} is the energy dependence of the K'-valley filtered transmission  of on the $\alpha$-strain parameter, for $\gamma=0.83$ and $\alpha$ varying from  $2.2$ to $3.4\%$. Small changes in the curve are observed, mainly in the intensity, with the filter energy window varying from  $14$ to $21 meV$, but with almost the same $L_Z/b$ window, being a bit larger for larger $\alpha$. This analysis is valid for $\alpha <3.5\%$. The valley filter energy window starts to decrease for higher strain values since other states from the second LL start to contribute to the transmission at the same energy window.  

\section{Conclusion}

Multi-terminal graphene systems are addressed as appropriate set-ups for valley filter and switches of valley polarized current directions. Using the mode-matching model based on the Green's functions formalism, we showed that strain enhances the valley filtering processes in graphene multi-terminal configurations. For extended folds in the longitudinal directions, currents flowing from left to top/bottom leads are valley polarized, with larger energy windows for higher strain intensity. An external magnetic field improves the deformation effects on the valley filter processes, avoiding undesirable disorder outcomes, and generating valley filter energy windows for lower strain intensities. Adding the magnetic field also switches the valley filter direction in some energy windows, with valley filtered transmission from the left to the right contacts.  We showed a diagram analyzing the interplay between the confinement parameters that maps the valley filter energy window, including the dependence on the strain intensity $\alpha$, the strain spread $L_Z/b$, and the ratio between the magnetic length and the deformation extension, $\gamma$. We hope this analysis motivates the development of other works with experimental measurements and applications of valley polarized electronic currents.


\section*{Acknowledgments}

This work was financially supported by the Brazilian Agencies CAPES, CNPq, and FAPERJ under the grant E-26/202.567/2019, and the INCT de Nanomateriais de Carbono. VT thanks the grant E-26/202.322/2018 from FAPERJ. We are grateful to Nancy Sandler for helpful discussions.

\bibliographystyle{elsarticle-num}
\bibliography{ref}

\end{document}